\documentclass[elsart12]{elsart}
\usepackage{epsfig}
\begin{document}
\begin{frontmatter}

\title{Absorption in periodic layered structures}

\author{Alexander Moroz\thanksref{www}}
\address{I. Institut f\"{u}r Theoretische Physik,  Jungiusstrasse 9, 
Universit\"{a}t Hamburg, D-20355 Hamburg, Germany}

\author{Adriaan Tip\thanksref{www}} 
\address{FOM-Instituut voor  Atoom- en Molecuulfysica, Kruislaan 407, 
1098 SJ Amsterdam,  The Netherlands}

\author{Jean-Michel Combes}
\address{D\'{e}partement de Math\'{e}matiques,
Universit\'{e} de Toulon et du Var, F-83130 La Garde \\
Centre de Physique Th\'{e}orique, CNRS-Luminy, Case 907,
F-13288 Marseille Cedex 9, France}

\thanks[www]{http://www.amolf.nl/research/photonic\_materials\_theory}

\begin{abstract}
Photonic band structure of metal-dielectric and semiconductor-dielectric
layered structures are studied  in the presence of a strong absorption.
It is shown that absorption can enlarge some gaps by as much as $50\%$.
\end{abstract}

\end{frontmatter}

\section{Introduction}
There has been a growing interest in photonic crystals, i. e.,  
structures with a periodically modulated  dielectric constant.  
The latter open new ways of manipulating 
electromagnetic wave emission and propagation processes  \cite{By,Y}.
In fact, there is a  common belief that, in the near future, 
photonic crystals systems will  allow us to perform many  functions 
with light that ordinary crystals do with electrons \cite{Y}.
They also promise to  become a  laboratory for testing fundamental 
processes involving interactions of radiation with matter under 
novel conditions.

The basic principle behind all that is an analogy  
to an electron moving in a periodic potential which suggests that
certain photon frequencies in a photonic crystal can also 
become forbidden,  independent of photon polarization and 
the direction of propagation -  a complete photonic bandgap  (CPBG). 
However, unlike in an ideal single-electron picture, a photon always 
faces the possibility of being absorbed and it is, 
{\em a priori\/} not clear if  
theoretical proposals for manufacturing  a CPBG 
and which are obtained in the absence of 
absorption  \cite{HCS,AlM}  can ever be realized experimentally.
Naively, one expects that moderate absorption will only cause
a slight perturbation of the band structure as confirmed in 
Ref.  \cite{KHa}. Therefore, the neglect of absorption in 
band structure calculations
seem to be justified for purely dielectric photonic structures  
\cite{YGL}. Unfortunately, such structures exhibit only a 
limited dielectric contrast
which is insufficient to open a CPBG below infrared 
wavelengths \cite{SHI}.
A recent proposal  \cite{AlM}  to obtain such a CPBG involves
metallo-dielectric structures. Since metals are usually
rather absorbing, this again raises the question about the 
effect of absorption on band
structure.
Here we mention in passing that systems with gain, i.e., where the 
imaginary part of the
dielectric function $\varepsilon$ has the opposite sign,  are 
important for understanding
laser action in new types of lasers involving 
photonic crystals \cite{JS}.

Although the study of absorption in photonic structures is of
utmost importance, so far no rigorous results exist about 
fundamental matters such as the meaning of band structure and 
band gaps for absorptive systems. Only recently,
using analytic continuation techniques, it has been shown that 
absorption turns bands into
resonances  in the lower-half complex plane  \cite{TMC}.
Literature on  absorption in photonic crystals is rather sparse 
and, except for a recent article by 
Yannopapas, Modinos, and Stefanou \cite{YMS}, only covers the cases 
of small absorption  \cite{KHa}  and small filling
fraction ($f_a\leq 1\%$) of the absorptive component  \cite{SSC}.

\section{Theory}
In order to investigate numerically the effect of considerable 
absorption on the band structure of photonic crystals, we started 
with studying a one-dimensional model
consisting of a periodically  layered (or stratified) medium, described 
by  the dielectric function
\begin{equation}
\varepsilon({\bf x})=\left\{
\begin{array}{cl}
\varepsilon_s, & {\bf x}\in (na-r_s/2,na+r_s/2)\\ 
\varepsilon_h, & {\bf x}\not\in  (na-r_s,na+r_s).
\end{array}\right.
\end{equation}
Here  $a$ is the lattice constant, i.e., the length of the 
primitive cell, $n$ is an integer, and $r_s < a/2$.
Propagation of light  at normal incidence in such a stratified medium
is described by  the one-dimensional  periodic Helmholtz equation
\begin{equation}
\left[\triangle +\varepsilon({\bf x}) \frac{\omega^2}{c^2}\right]\,\psi({\bf x})=0,
\label{1dhe}
\end{equation}
where $\triangle$ is the Laplacian, $\omega$ the angular frequency, 
$c$  the speed of light in vacuum and $\psi$ an eigenfunction
to be calculated. Another physical situation to which 
the one-dimensional  periodic Helmholtz equation applies (\ref{1dhe})
are  grating-like structures  \cite{RK},  the so called
one-dimensional (1d) photonic crystals \cite{For}, 
and  a number of different physical situations involving propagation of 
acoustic and  elastic waves \cite{MF}.

Below a scatterer will be a region with  
$\varepsilon({\bf x})=\varepsilon_s$ neighboured by regions with 
$\varepsilon({\bf x})=\varepsilon_h$. 
For notational simplicity we also set $\omega=\omega/c$.
In the absence of absorption, the dispersion relation 
and the band structure are determined by the KKR equation, which in one dimension
is identical to the Kronig-Penney equation \cite{KP}
\begin{equation}
s= \cos(\sigma a + \eta_0 +\eta_1) /\cos (\eta_0 -\eta_1) =\cos {\bf k}a, 
\label{kkreq}
\end{equation}
where ${\bf k}$ is the Bloch momentum and  $\eta_l$ is a phase shift \cite{But}. 
The latter is determined by relations
\begin{eqnarray}
\cos\eta_0 &=& C_0\left[\cos \sigma r_s\cos\rho r_s +
                      p\,\sin\sigma r_s \sin\rho r_s
\right],
 \nonumber  \\
\sin\eta_0 &=& C_0\left[-\sin \sigma r_s\cos\rho r_s +
                   p \,\cos\sigma r_s \sin\rho r_s
\right],
  \nonumber \\
\cos\eta_1 &=& C_1\left[\sin \sigma r_s\sin\rho r_s +
                     p \,\cos\sigma r_s \cos\rho r_s
\right],
  \nonumber \\
\sin\eta_1 &=& C_1\left[\cos \sigma r_s\sin\rho r_s -
                    p\,\sin\sigma r_s \cos\rho r_s
\right],
  \nonumber \\
C_l &=& \left\{
\begin{array}{cc}
\left[1 + (p^2-1)\,\sin^2\rho r_s\right]^{-1/2}, & l=0,\\
\left[ 1 + (p^2-1)\,\cos^2\rho r_s\right]^{-1/2}, & l=1,
\end{array}\right.
 \nonumber 
\end{eqnarray}
where  $\rho=\omega \sqrt{\varepsilon_s}$,  $\sigma=\omega\sqrt{\varepsilon_h}$,
and  $p=\sqrt{\varepsilon_s/\varepsilon_h}$ \cite{But}.

Equation  (\ref{kkreq}) still holds for  absorbing scatterers but 
the (frequency-dependent)
phase shifts $\eta_l$ are now complex. Consequently,  $s$  is complex and 
Eq. (\ref{kkreq}) has no real eigenvalues. In accordance with \cite{TMC}, 
the eigenvalues
turn into resonances in the  lower complex half-plane. 
\begin{figure}
\begin{center}
\epsfig{file=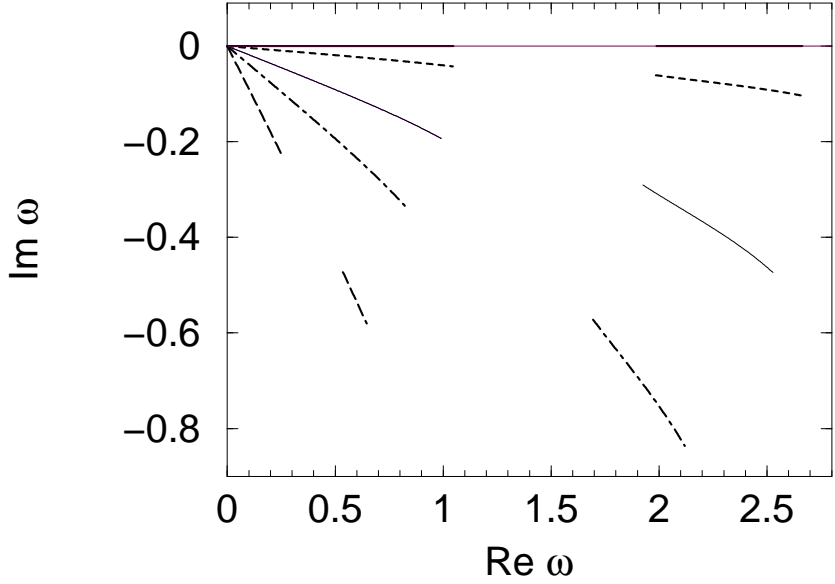,width=8cm,clip=0,angle=-90}
\end{center}
\caption{When absorption is increased, bands of resonances 
in the complex energy plane are at larger and larger angle with
respect to the real positive axis. First two complex bands are shown
for absorptive filling fraction $f_a=40\%$, Re $\varepsilon_s=12$,
and  $\varepsilon_h=1$
as Im $\varepsilon_s$ increases from $0$ to $1$, $5$, $12$, and $120$.}
\end{figure}
Some 
characteristic features
of the complex bands of resonances can already be seen in the 
long-wavelength limit ($k\ll 1$),
when Eq. (\ref{kkreq}) reduces to
\begin{equation}
\varepsilon_{e\!f\!f}\omega^2 \approx  k^2.
\end{equation}
Here $\varepsilon_{e\!f\!f}$ is the {\em effective dielectric constant\/}.
If we denote by $f_a$ the absorptive material filling  fraction, then
\begin{equation}
\varepsilon_{e\!f\!f}=(1-f_a) \varepsilon_h  + f_a \varepsilon_s,
\label{efdiel}
\end{equation}
i.e.,  $\varepsilon_{e\!f\!f}$ is nothing but  the volume 
averaged dielectric constant. In the long-wavelength limit, the 
first complex band contains an angle $\theta$
with the real positive axis which is determined by the relation 
\begin{equation}
\sin 2\theta = -  \frac{ \mbox{Im}\, 
(\varepsilon_s)}{|\varepsilon_{e\!f\!f}|}\, f_a .
\end{equation}
In the limit $ \mbox{Im}\, (\varepsilon_s)\rightarrow\infty$,
$\sin 2\theta = -1$
and hence $\theta\rightarrow  -\pi/4$. For 
$\mbox{Re}\,(\varepsilon_{s})\gg \varepsilon_h$
and $\mbox{Im}\,(\varepsilon_{s})\gg \varepsilon_h$ one has
\begin{equation}
\sin 2\theta \approx  -  \frac{ \mbox{Im}\, (\varepsilon_s)}{|\varepsilon_{s}|}
\end{equation}
down to such $f_a$  that 
$f_a\mbox{Re}\,(\varepsilon_{s})\gg \varepsilon_h$ and 
$f_a\mbox{Im}\,(\varepsilon_{s})\gg \varepsilon_h$.
In this range of $f_a$, $\theta$ exhibits only a very weak dependence on $f_a$.

\section{Numerical results}
Surprisingly enough, as shown in Fig. 1,  exact numerical calculations  
demonstrate that  the angle $\theta=\theta(\omega)$ 
remains very close to its  value established in the long-wavelength limit
often  till  the upper edge of the second 
complex band of resonances. 
We determined  the  resonances numerically for 
an ensemble of  1d systems with Im $\varepsilon_s$ varied between 
$0$ and $120$ for
 fixed  $\varepsilon_h=1$ and Re $\varepsilon_s=12$.
We investigated how absorption alters the relative gap width
$g_w=\Delta\omega/\omega_c$, where $\Delta\omega$ and $\omega_c$
are the gap width and midgap frequency of the real part of
$\omega$, respectively.
The following lessons can be learned from an investigation of the first 
few bands of complex resonances  for  a representative sample of 
such 1d absorptive  
systems:

\begin{enumerate}

\item if (Im $\varepsilon_s$)/(Re $\varepsilon_s) \leq 0.1$, the effect 
of absorption
   on the band gap is negligible (both in shifting
   the gap edges and in changing gap width) (see Figs. 2 and 3)

\item  Re $\omega({\bf k})$ remains a monotonically increasing (decreasing)
function of the Bloch momentum for an  odd (even) band
(when labelled from the lowest one)

\item absorption pushes band edges down  (see Fig. 2)

\item if Im $\varepsilon_s$ increases, the corresponding $g_{w}$
tend to saturate rapidly beyond  Im $\varepsilon_s\approx $ Re $\varepsilon_s$
 (see Fig. 3)

\item  the dependence of the relative gap width $g_{w}$ on Im $\varepsilon_s$ is 
monotonic up to  Im $\varepsilon_s\approx 3\times $ Re $\varepsilon_s$
 (see Fig. 3)

\item  absorption mostly increases the relative gap width $g_{w}$,
although in some rare cases it can also lead to a decrease of  $g_{w}$

\item  the largest effect was an increase of the $g_{w}$  by  $50\%$   (see Fig. 3)
and closing of the 2nd gap which, however, reappeared again as
Im $\varepsilon_s$ increased sufficiently enough

\item Even if the absorptive material filling  fraction $f_a$
   within the unit cell is $99\%$ and Im $\varepsilon_s$
   is comparable to  Re $\varepsilon_s$, the imaginary part
   of the Bloch  frequency eigenvalues is
   $\leq (\mbox{Im}\, \varepsilon_s)/15$.
   Actually, it has been quite surprising to
   find out that the imaginary part
   of the frequency eigenvalues can increase when
   the absorptive material filling  fraction
   within the unit cell is lowered!

\end{enumerate}
\begin{figure}
\begin{center}
\epsfig{file=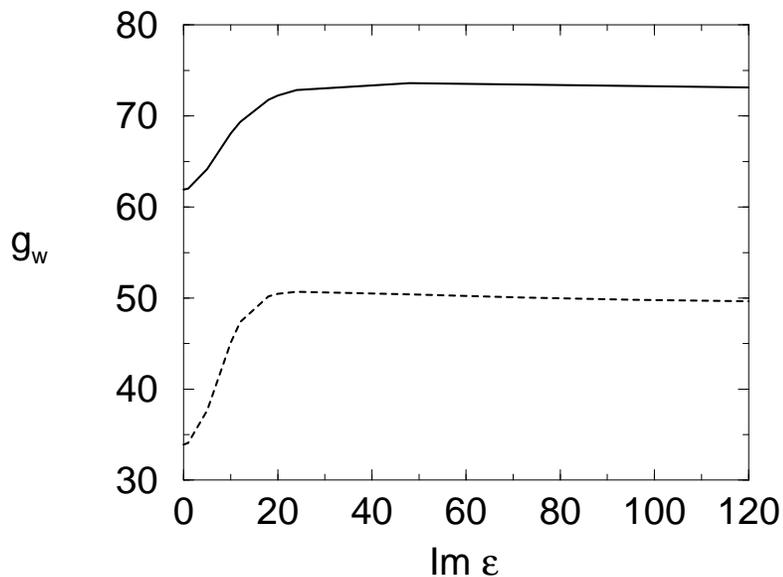,width=8cm,clip=0,angle=-90}
\end{center}
\caption{As absorption increases, the band edges are 
pushed lower and lower. Here it is demonstrated for 
band edges of the real part  of the first three bands of complex resonances
as a function of Im $\varepsilon_s$. Remaining  parameters 
are as in Fig. 1.
}
\end{figure}
\begin{figure}
\begin{center}
\epsfig{file=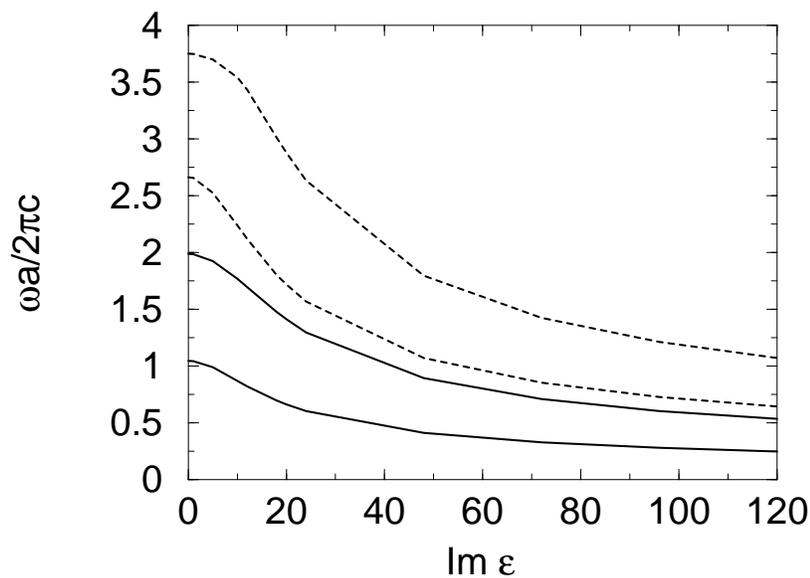,width=8cm,clip=0,angle=-90}
\end{center}
\caption{
The relative gap width of a gap in the real part
of bands of  complex resonances as a function of Im $\varepsilon_s$. 
Remaining parameters are as in Figs. 1 and 2.
}
\end{figure}

It is noteworthy to say that a band gap in Re $\omega$ does not necessarily 
imply a very high reflectivity. A study of finite layered structures
reveals that in the presence of a strong absorption the reflectivity
at the gap center can, for instance, saturate at $60\%$ or less, depending
on a single bilayer absorption. The latter yields a lower bound on the
absorption of a composite structure.
The reflectivity at the gap center is usually more than twice that for frequencies
within a band, almost exclusively as a result of decreased absorption 
within the gap which approaches that of a single bilayer. In both cases 
transmissions are small.
If $T$ denotes the transmission for frequencies within a band, 
then the transmission within a gap is typically $T^2$.

A typical system to which our calculation should apply is
a polymer-semiconductor or polymer-insulator layered structure
where the polymer absorption dominates.
Our model calculations can also apply to a  semiconductor-semiconductor or
semiconductor-insulator
heterostructure in which one component is doped by an increasing 
amount of an absorptive dye. Our choice of $\varepsilon_s$ can also be 
regarded as that of  a metal, or an artificial  metal \cite{PHS}, 
for frequencies above the plasma wavelength.

\section{Conclusions}
To our best knowledge, our work is the first one to deal with the case 
of  the photonic band structure in the presence of a strong absorption, when both
$\mbox{Im}\, \varepsilon_s$ and $f_a$ are large
($\mbox{Im}\, \varepsilon_s\sim  10\times \mbox{Re}\, \varepsilon_s =120$ 
and $f_a\sim 99\%$) (cf. \cite{YMS,KHa,SSC}).
Such a study of strong absorption in photonic structures is of
utmost importance, both for understanding the behaviour of  recently proposed  
metallo-dielectric structures with CPBG's below infrared wavelengths  \cite{AlM}
and for
new types of lasers involving photonic crystals \cite{JS}.
More realistic calculations involving dispersion of the dielectric constant
and including transmission and reflection properties of finite layered
structures  will appear elsewhere.

This work is part of the research program by  the Stichting voor 
Fundamenteel Onderzoek der Materie  (Foundation for Fundamental 
Research on Matter) which  was made possible by financial support from the 
Nederlandse Organisatie voor Wetenschappelijk Onderzoek 
(Netherlands Organization for Scientific Research).


\newpage


\begin{thebibliography}{99}

\bibitem{By}V. P. Bykov, Sov. Phys. JETP {\bf 35}, 269 (1972); 
Sov. J. Quant. Electron. {\bf 4}, 861 (1975).

\bibitem{Y}E. Yablonovitch,
Phys. Rev. Lett. {\bf 58}, 2059 (1987).

\bibitem{HCS}K. M. Ho, C. T. Chan and C. M. Soukoulis,
Phys. Rev. Lett. {\bf 65}, 3152 (1990);
M. Plihal, A. Shambrook, A. A. Maradudin, and P. Sheng,
Optics Commun. {\bf 80}, 199 (1991).

\bibitem{AlM}A. Moroz, Phys. Rev. Lett. {\bf 83}, 5274 (1999);
Photonic crystals of coated metallic spheres, to appear in Europhys. Lett.;
H. van der Lem and A. Moroz,  Towards two-dimensional complete 
photonic-band-gap structures below infrared wavelengths, 
to appear in J.  Opt. A: Pure Appl. Opt. 

\bibitem{YMS}V. Yannopapas, A. Modinos, and N. Stefanou, 
Phys. Rev. B {\bf 60}, 5359 (1999).

\bibitem{KHa}A. A. Krokhin and P. Halevi, Phys. Rev. B {\bf 53}, 1205 (1996).

\bibitem{SSC}M. M. Sigalas, C. M. Soukoulis, C. T. Chan, and  K. M. Ho, 
Phys. Rev. B {\bf 49}, 11 080 (1994);
V. Kuzmiak and  A. A. Maradudin, Phys. Rev. B {\bf 55}, 7427 (1997);
L.-M. Li, Z.-Q. Zhang, and X. Zhang, Phys. Rev. B {\bf 58}, 15589 (1998).

\bibitem{YGL}E. Yablonovitch, T. J. Gmitter, and K. M. Leung,
Phys. Rev. Lett. {\bf 67}, 2295 (1991).

\bibitem{SHI}H. S. S\"{o}z\"{u}er, J. W. Haus, and R. Inguva,
Phys. Rev. B {\bf 45}, 13962 (1992);
A. Barra, D. Cassagne, and C. Jouanin,
Appl. Phys. Lett. 72, 627 (1998);
R. Biswas, M. M. Sigalas, G. Subramania, and K.-M. Ho, 
Phys. Rev. B {\bf 57}, 3701 (1998); 
A. Moroz and C. Sommers, 
J. Phys.: Condens. Matter {\bf 11}, 997 (1999).

\bibitem{JS}X. Jiang and C. M. Soukoulis, Physical Review B {\bf 59},  6159 (1999). 

\bibitem{TMC}A. Tip, A. Moroz, and J.-M. Combes,
Bloch decomposition and band structure for absorptive photonic
crystals, submitted for publication.

\bibitem{RK}P. Rigby and T. F. Krauss, Nature {\bf 390}, 125 (1997).

\bibitem{For}J. S. Foresi et al.,  Nature {\bf 390}, 143 (1997).

\bibitem{MF}P. M. Morse and H. Feshbach, {\em Methods of Theoretical 
Physics\/}
(Mc-Graw Hill, New York, 1953), Part II, Secs. XI, XIII.

\bibitem{KP}R. de L. Kronig and W. G. Penney, Proc. Roy. Soc. A
{\bf 20}, 499 (1931).

\bibitem{But}W. H. Butler, Phys. Rev. B {\bf 14}, 468 (1976).

\bibitem{PHS}J. B. Pendry, A. J. Holden, W. J. Stewart, 
and I. Youngs, Phys. Rev. Lett. {\bf 76}, 4773 (1996).

\end{thebibliography}
\end{document}